\documentclass[aps,prc,twocolumn,showpacs,superscriptaddress,footinbib]{revtex4-1}

\usepackage{graphicx}
\usepackage{dcolumn}
\usepackage{bm}
\usepackage{enumerate}

\usepackage[normalem]{ulem}  
\usepackage[dvips]{color}

\begin{document}


\title{Influence of the density of states on the odd-even staggering in the charge distribution of the emitted fragments}

\author{N.L. Calleya}
\affiliation{Instituto de F\'\i sica, Universidade Federal do Rio Grande do Sul,\\
Av. Bento Gon\c calves 9500, CP 15051, 91501-970, Porto Alegre, Brazil}
\author{S.R. Souza}
\affiliation{Instituto de F\'\i sica, Universidade Federal do Rio de Janeiro Cidade Universit\'aria, \\CP 68528, 21941-972, Rio de Janeiro, Brazil}
\affiliation{Instituto de F\'\i sica, Universidade Federal do Rio Grande do Sul,\\
Av. Bento Gon\c calves 9500, CP 15051, 91501-970, Porto Alegre, Brazil}
\author{B.V. Carlson}
\affiliation{Departamento de F\'\i sica, Instituto Tecnol\'ogico de Aeron\'autica-CTA, 12228-900, S\~ao Jos\'e dos Campos, Brazil}
\author{R. Donangelo}
\affiliation{Instituto de F\'\i sica, Universidade Federal do Rio de Janeiro Cidade Universit\'aria, \\CP 68528, 21941-972, Rio de Janeiro, Brazil}
\affiliation{Instituto de F\'\i sica, Facultad de Ingenier\'\i a, Universidad de la Rep\'ublica, Julio Herrera y Reissig 565, 11.300 Montevideo, Uruguay}
\author{W.G. Lynch}
\affiliation{National Superconducting Cyclotron Laboratory and Department of Physics and Astronomy Department,\\ Michigan State University, East Lansing, Michigan 48824, USA}
\author{M.B. Tsang}
\affiliation{National Superconducting Cyclotron Laboratory and Department of Physics and Astronomy Department,\\ Michigan State University, East Lansing, Michigan 48824, USA}
\author{J. R. Winkelbauer}
\affiliation{National Superconducting Cyclotron Laboratory and Department of Physics and Astronomy Department,\\ Michigan State University, East Lansing, Michigan 48824, USA}

\date{\today}

\begin{abstract}
The fragmentation of thermalized sources is studied using a version of the Statistical Multifragmentation Model which employs state densities that take the pairing gap in the nuclear levels into account.
Attention is focused on the properties of the charge distributions observed in the breakup of the source.
Since the microcanonical version of the model used in this study provides the primary fragment excitation energy distribution, one may 
correlate the reduction of the odd-even staggering in the charge distribution with the increasing occupation of high energy states.
Thus, in the framework of this model, such staggering tends to disappear as a function of the total excitation energy of the source, although the energy per particle may be small for large systems.
We also find that, although the deexcitation of the primary fragments should, in principle, blur these odd-even effects as the fragments follow their decay chains, the consistent treatment of pairing may significantly enhance these staggering effects on the final yields.
In the framework of this model, we find that odd-even effects in the charge distributions should be observed in the fragmentation of relatively light systems at very low excitation energies.
Our results also suggest that the odd-even staggering may provide useful information on the nuclear state density.
\end{abstract}

\maketitle

\begin{section}{Introduction}
\label{sect:introduction}
Size and isotopic correlations of fragments  produced in nuclear reactions provide important information both on the reaction mecanisms \cite{Moretto1993,Bondorf1995,PochodzallaReview1997,TanIsoEOS} and the nuclear properties \cite{Moretto1993,Bondorf1995,PochodzallaReview1997,TanIsoEOS,riseandfall,Peaslee1994,reviewSubal2001,BettyPhysRep2005}.
For instance, the nuclear isoscaling \cite{isoscaling1,isoscaling2} turned out to hold information on the qualitative shape of the nuclear caloric curve \cite{isocc} and on the symmetry energy \cite{isoscaling3},
although further investigations revealed that  care must be taken in drawing conclusions based on such analyses \cite{isoMassFormula2008,isoSMMTF,isotemp,isoscalingDasGupta2009,finiteSizeEffects2012}.

Odd-even effects on the fragment charge distributions produced in different reactions have been recently reported in the literature \cite{StaggeringRicciardi2005,StaggeringDagostino2011,StaggeringCasini2012,Staggering2013,StaggeringCasini2013}.
Analyses based on the fragmentation of the quasi-projectile have been made at relativistic energies \cite{StaggeringRicciardi2005} as well as at much lower bombarding energies \cite{Staggering2013}.
In both cases, clear odd-even effects have been observed in the fragment size distribution.
This is  surprising, to a certain extent, since the pairing gap should quickly vanish as the system is heated up \cite{PairingRingShuck1989,pairingGoriely1996}.
The data reported in Ref.\  \cite{StaggeringRicciardi2005} has been analyzed using an abrasion-evaporation model and the odd-even effects were attributed to the late stages of the evaporation process, during which the system is relatively cool.
In Ref.\ \cite{Staggering2013}, it is demonstrated that the deexcitation of the primary hot fragments plays a very important role.
Indeed, it was found that the adopted deexcitation process leads to the appearance of staggering effects in charge correlations of fragments with odd neutron excess,
not observed in the primary fragments.
Similar conclusions were also drawn in Ref.\ \cite {StaggeringDagostino2011}, where it was  suggested that staggering should occur at low excitation energies.
The study of central and semi-peripheral collisions carried out in Ref.\ \cite{StaggeringCasini2012} shows that important odd-even effects are observed for fragments with $Z<15$ and, in the case of peripheral collisions, they can be observed up to $Z=40$.
On the other hand, other experimental results \cite{StaggeringCasini2013} reveal that these effects rapidly decrease as the fragment size increases.

In this work, we investigate the odd-even staggering using the version of the Statistical Multifragmentation Model (SMM) presented in Ref.\ \cite{smmde2013}.
In this implementation, the deexcitation of the primary fragments is treated using a generalization of the Fermi breakup model (GFBM) \cite{fbk2012}, in which the emitted fragments are excited.
As was demonstrated in Ref.\ \cite{fbk2012}, this is equivalent to the standard version of SMM if the same ingredients are employed in both treatments.
It therefore allows one to investigate the role played by the pairing energy at the different stages of the process if it is consistently taken into account in the model.

We thus start, in Sec.\ \ref{sect:theory}, by reviewing the main features of the treatment presented in Ref.\ \cite{smmde2013} and discuss the modifications of the model needed to include pairing effects in the nuclear state density.
The predictions of the model are presented and discussed in Sec.\ \ref{sect:results}.
Concluding remarks are drawn in  Sec.\ \ref{sect:conclusions}.

\end{section}

\begin{section}{Theoretical framework}
\label{sect:theory}
The SMM \cite{smm1,smm2,smm4} assumes that an equilibrated source of mass and atomic numbers $A_0$ and $Z_0$, respectively, breaks up simultaneously after having expanded to a breakup volume $V=(1+\chi) V_0$, where $V_0$ is the volume corresponding to the normal nuclear density.
We use $\chi=2$ in this work.
Besides strict mass and charge conservation, each fragmentation mode must fulfill the energy conservation constraint:

\begin{equation}
E^*-B_{A_0,Z_0}=C_c\frac{Z_0^2}{A_0^{1/3}}\frac{1}{(1+\chi)^{1/3}}+\sum_{\{A,Z\}}n_{A,Z}E_{A,Z}\;,
\label{eq:econst}
\end{equation}

\noindent
where $E^*$ is the total excitation energy of the source and $B_{A_0,Z_0}$ denotes its binding energy.
Except for fragments with $A\le 4$, for which empirical values are adopted, we use the mass formula developed in Ref.\ \cite{ISMMmass}:

\begin{equation}
B_{A,Z}=C_vA-C_sA^{2/3}-C_c\frac{Z^2}{A^{1/3}}+C_d\frac{Z^2}{A}+\delta_{A,Z}A^{-1/2}\;,
\label{eq:be}
\end{equation}

\noindent
where

\begin{equation}
C_i=a_i\left[1-k\left(\frac{A-2Z}{A}\right)^2\right]\;.
\end{equation}

\noindent
and $i=v,s$ denotes the volume and surface terms, respectively.
The last term in Eq.\ (\ref{eq:be}) is the usual pairing contribution to the binding energy:

\begin{equation}
\delta_{A,Z}=\frac{1}{2}\left[(-1)^{A-Z}+(-1)^Z\right]C_p\;.
\label{eq:delta}
\end{equation}

\noindent
The numerical values of all the parameters are listed in Ref.\ \cite{ISMMmass}.

The Coulomb interaction between the fragments is taken into account in the framework of the Wigner-Seitz approximation \cite{smm1,WignerSeitz}:
\begin{eqnarray}
E_{\rm Coul} &=&C_c\frac{Z_0^2}{A_0^{1/3}}\frac{1}{(1+\chi)^{1/3}}\nonumber\\
&+&C_c\sum_{A,Z}n_{A,Z}\frac{Z^2}{A^{1/3}}\left[1-\frac{1}{(1+\chi)^{1/3}}\right]\;,
\label{eq:ecws}
\end{eqnarray}

\noindent
where $n_{A,Z}$ denotes the multiplicity of a species $(A,Z)$. 
The contribution associated with the homogeneous sphere of volume $V$ is explicitly written on the right hand side of Eq.\ (\ref{eq:econst}), whereas the other terms are contained in $E_{A,Z}$, which reads:

\begin{equation}
E_{A,Z}=-B_{A,Z}+\epsilon^*_{A,Z}-C_c\frac{Z^2}{A^{1/3}}\frac{1}{(1+\chi)^{1/3}}
+E_{A,Z}^{\rm trans}\;,
\label{eq:eaz}
\end{equation}

\noindent
where $\epsilon^*_{A,Z}$ represents the excitation energy of the fragment and $E_{A,Z}^{\rm trans}$ is its translational energy.

In order to employ the efficient recursion formulae developed in Ref.\ \cite{PrattDasGupta2000}, Eq. (\ref{eq:econst}) is conveniently rewritten as:

\begin{equation}
Q\Delta_Q\equiv E^*-B^c_{A_0,Z_0}=\Delta_Q\sum_{\alpha,q_\alpha} q_\alpha n_{\alpha,q_\alpha} \;,
\label{eq:qconst}
\end{equation}

\noindent
so that $Q$ is an integer number. In this way, the energy is discretized and the parameter $\Delta_Q$ controls the granularity of the discretization.
The quantity $B_{A,Z}^c$ corresponds to

\begin{equation}
B^c_{A,Z}\equiv B_{A,Z}+C_c\frac{Z^2}{ A^{1/3}}\frac{1}{(1+\chi)^{1/3}}
\label{eq:bcaz}
\end{equation}

\noindent
whereas

\begin{equation}
q_{A,Z}\Delta_Q\equiv -B^c_{A,Z}+\epsilon^*_{A,Z}+E_{A,Z}^{\rm trans}\;.
\label{eq:qaz}
\end{equation}

\noindent
The sum over $\alpha$ is carried out through all the isotopic species whereas that over $q_\alpha$ must be consistent with energy conservation, as stated in Eq.\ (\ref{eq:qconst}).
Following Refs.\ \cite{PrattDasGupta2000} and \cite{smmde2013}, the average multiplicity of a species $(a,z)$, with energy $q\Delta_Q$, is given by:

\begin{equation}
\overline{n}_{a,z,q}=\frac{\omega_{a,z,q}}{\Omega_{A_0,Z_0,Q}}\Omega_{A_0-a,Z_0-z,Q-q}\;.
\label{eq:nazq}
\end{equation}

\noindent
The statistical weight $\Omega_{A,Z,q}$ is calculated through the following recurrence relation:

\begin{equation}
\Omega_{A,Z,Q}=\sum_{\alpha,q_\alpha}\frac{a_\alpha}{A}\omega_{a_\alpha,z_\alpha,q_{\alpha}}\Omega_{A-a_\alpha,Z-z_\alpha,Q-q_{\alpha}}\
\label{eq:oazq}
\end{equation}

\noindent
and $\omega_{A,Z,q}$ is obtained by folding the number of states associated with the kinetic motion with that corresponding to  the internal degrees of freedom:

\begin{equation}
w_{A,Z,q}=\gamma_A\int_0^{\epsilon_{A,Z,q}}\,dK\;\sqrt{K}\rho_{A,Z}(\epsilon_{A,Z,q}-K)\;,
\label{eq:wazq}
\end{equation}

\noindent
where

\begin{equation}
\gamma_A=\Delta_Q \frac{V_f (2m_n A)^{3/2}}{4\pi^2\hbar^3}\;,
\label{eq:gamma}
\end{equation}

\noindent
$\epsilon_{A,Z,q}\equiv {\,q\Delta_Q+B^c_{A,Z}}$, $V_f=\chi V_0$ represents the free volume, $m_n$ is the nucleon mass, and $\rho_{A,Z}(\epsilon^*)$ is the density of the internal states of the nucleus $(A,Z)$ with excitation energy $\epsilon^*$.
Thus, once the state density is specified, the above relations allow one to calculate the statistical properties of the system.

The final fragment distribution is obtained by applying this treatment successively for each fragment until they have decayed to the final states, as described in Refs.\ \cite{fbk2012,smmde2013}.
More specifically, each species $(A,Z)$ contributes to the yields of $(a,z)$ which add up to:

\begin{equation}
\overline{n}_{a,z,q}\rightarrow\overline{n}_{a,z,q}+\frac{\overline{n}'_{a,z,q}}{1-\overline{n}'_{A,Z,q_0}}\times\overline{n}_{A,Z,q_0}\;,\;\;\; a < A\;.
\label{eq:yrenrom}
\end{equation} 

\noindent
where

\begin{equation}
\overline{n}'_{a,z,q}=\frac{\omega_{a,z,q}}{\Omega_{A,Z,q_0}}\Omega_{A-a,Z-z,q_0-q}\;.
\label{eq:nazqp}
\end{equation}

\noindent
Thus, by starting from the heaviest species down to the lightest fragment, one generates the final distribution.

The above relationships clearly show that the density of states plays a key role in the different stages of the process.
The traditional SMM model employs \cite{ISMMlong}:

\begin{equation}
\rho_{A,Z}(\epsilon^*)=\rho_{\rm SMM}(\epsilon^*)=\rho_{\rm FG}(\epsilon^*)e^{-b_{\rm SMM}(a_{\rm SMM}\epsilon^*)^{3/2}}
\label{eq:rhoSMM}
\end{equation}

\noindent
with

\begin{equation}
\rho_{\rm FG}(\epsilon^*)=\frac{a_{\rm SMM}}{\sqrt{4\pi}{(a_{\rm SMM}\epsilon^*)}^{3/4}}\exp(2\sqrt{a_{\rm SMM}\epsilon^*})
\label{eq:rhofg}
\end{equation}

\noindent
and

\begin{equation}
a_{\rm SMM}=\frac{A}{\epsilon_0}+\frac{5}{2}\beta_0\frac{A^{2/3}}{T_c^2}\;,
\label{eq:asmm}
\end{equation}

\noindent
where $\epsilon_0=16.0$ MeV, $\beta_0=18.0$ MeV, and $T_c=18.0$ MeV.
The other parameters read $b_{\rm SMM}=0.07 A^{-\tau}$ and $\tau=1.82(1+A/4500)$, for $A>4$.
In the case of the alpha particles, we set $\beta_0=0$ and $b_{\rm SMM}=0.000848416$.
For the other light nuclei with $A<5$, which have no internal degrees of freedom, we use $\rho_{A,Z}(\epsilon^*)=g_{A,Z}\delta(\epsilon^*)$,
where $g_{A,Z}$ represents the empirical spin degeneracy factor.
To avoid numerical instabilities at very small excitation energies, in this work, we use $\rho_{A,Z}(\epsilon^*)=\rho_0e^{(\epsilon^*-U_x)/\tilde\tau}$ for $\epsilon^* < U_x$, where $U_x$ is defined below.
The parameters $\rho_0$ and $\tilde\tau$ are adjusted, for each species, in order to match the value and the first derivative of the density of states at $\epsilon^*=U_x$.

\begin{figure}[tbh]
\includegraphics[width=8.5cm,angle=0]{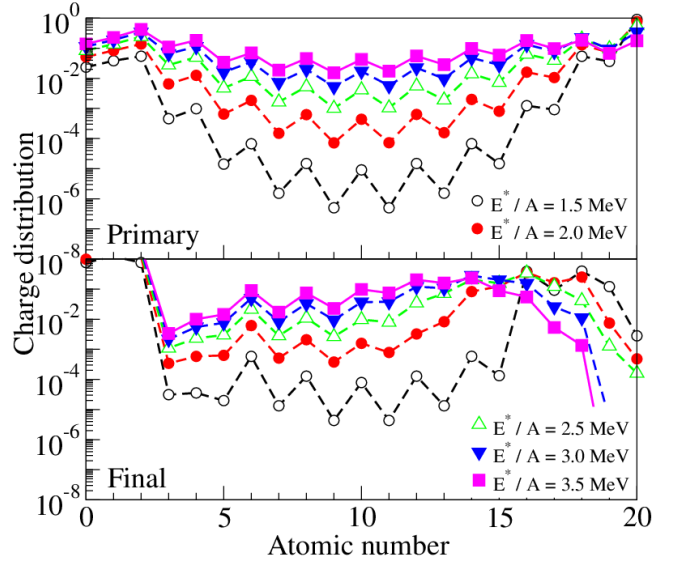}
\caption{\label{fig:smmpw} (Color online) Charge distribution from the breakup of the $^{40}$Ca nucleus at different excitation energies.
The top and bottom panels exhibit primary and final yields, respectively. The standard SMM state density, Eq.\ (\ref{eq:rhoSMM}), is used in all the cases. For details see the text.}
\end{figure}

Since this parameterization of the state density does not take pairing effects into account, this aspect is not consistently treated by the model.
For this reason, these effects appear even at high excitation energies.
This is illustrated in Fig.\ \ref{fig:smmpw} which shows the charge distribution obtained in the breakup of the $^{40}$Ca nucleus at different excitation energies.
The primary and final yields are respectively displayed at the top and bottom panels.
In both cases, odd-even staggering is clearly seen in the charge distribution which, in the framework of the model, can be explained only by the presence of the pairing term in the fragments' binding energy \cite{Staggering2013}, as is explicitly written in Eqs.\  (\ref{eq:be}) and (\ref{eq:delta}).
However, one should note that the breakup temperatures (see Eq.\ (\ref{eq:disc}) below) vary from $T= 3.8$ MeV for $E^*/A = 1.5$ MeV to $T=5.0$ MeV for $E^*/A=3.5$ MeV.
In this temperature range, pairing effects should not be so important \cite{PairingRingShuck1989,pairingGoriely1996}.

Empirical information on discrete states has been taken into account in the SMM version presented in Ref.\ \cite{ISMMlong}.
However, extremely exotic nuclei enter in the above recursion formulae, for which information on such states is very scarce or unavailable.
Furthermore, the quick growth of the number of states with the system size makes the implementation of this procedure extremely difficult, except in the case
of very light nuclei.
For this reason, in that work, analytical approximations have been used to supplement empirical information.

For simplicity, for all excitable nuclei, we adopt the parameterization proposed by Gilbert and Cameron \cite{levelDensityGilbertCameron1965}:

\begin{equation}
\label{eq:rhogc}
\rho_{\rm GC}(\epsilon^*)=
\cases{
\frac{\sqrt{2\pi}\sigma_0}{\tau}e^{(\epsilon^*-E_0)/\tau},&$\epsilon^* \le E_x$,\cr
\frac{\sqrt{\pi}}{12}\frac{e^{2\sqrt{a(\epsilon^*-\Delta)}}}{(\epsilon^*-\Delta)(a[\epsilon^*-\Delta])^{1/4}},& $\epsilon^* > E_x$,
}
\end{equation}

\noindent
where $\Delta$ is the pairing energy of the nucleus, $E_x=U_x+\Delta$, $U_x=2.5 +150/A$ (MeV), $\sigma_0^2=0.0888\sqrt{a(E_x-E_0)}A^{2/3}$,
$E_0=E_x-\tau\log[\tau\rho_2(E_x)]$,

\begin{equation}
\rho_2(E_x)=\frac{1}{12\sqrt{2}}\frac{1}{\sigma_0}\frac{e^{2\sqrt{a(E_x-\Delta)}}}{(E_x-\Delta)(a[E_x-\Delta])^{1/4}}\;,
\label{eq:rho2}
\end{equation}

\noindent
and

\begin{equation}
\frac{1}{\tau}=\sqrt{\frac{a}{E_x-\Delta}}-\frac{3}{2}\frac{1}{E_x-\Delta}\;.
\end{equation}

\begin{figure}[tbh]
\includegraphics[width=8.5cm,angle=0]{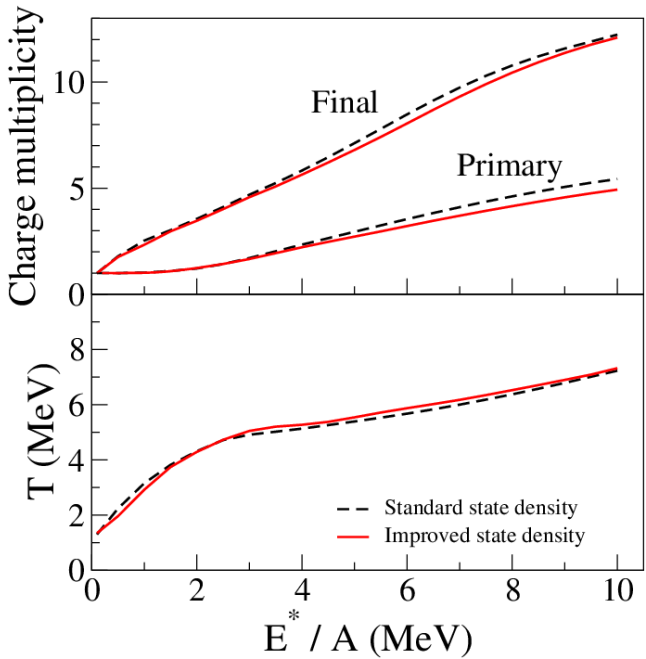}
\caption{\label{fig:aveobs} (Color online) Comparison of different observables, from  the breakup of the $^{40}$Ca nucleus, obtained with the standard state density given by Eq.\ (\ref{eq:rhoSMM})
and the improved formula Eq.\ (\ref{eq:rhom}).
For details see the text.}
\end{figure}

\noindent
For all nuclei, we use the level density parameter $a=A/8.0$~MeV$^{-1}$.

The low energy part of this state density takes into account the fact that, in this energy domain, collective modes are also excited, besides those associated with single particles states, so that the density of states
is enhanced compared to that of a Fermi gas due to this extra contribution.
For this reason, $\rho(\epsilon^*)$ increases as a function of $\exp({\epsilon^*})$ at low energies, instead of $\exp({2\sqrt{a\epsilon^*}})$ as it does at higher energies.
Its explicitly dependence on $\epsilon^*-\Delta$ for $E>E_x$, rather than on $\epsilon^*$, takes into account the fact that the nucleon pairs must break in order to excite single particle states.
In this way, the use of Eq.\ (\ref{eq:rhogc}) takes into account pairing effects in the excited states.

Despite this desirable feature, this density of states differs from that adopted in the standard SMM, which should lead to predictions being at odds with previous results.
Since we intend to preserve the properties of the model at high energies, we proceed as in Ref.\ \cite{ISMMlong} and gradually switch from $\rho_{\rm GC}$ to
$\rho_{\rm SMM}$,  so that we use:

\begin{equation}
\label{eq:rhom}
\rho(\epsilon^*)=
\rho_{\rm GC}(\epsilon^*) \left[1-f(x)\right]+\rho_{\rm SMM}(\epsilon^*-\Delta) f(x)\;.
\end{equation}

\noindent
There is freedom in choosing the function $f(x)$, as long as it leads to a smooth switch from the two expressions for the density of states.
We adopt $f(x)=[1+\tanh(x)]/2$, with $x=[\epsilon^*-E_x-(1/2)\Delta E]/\delta E$, $\Delta E/A=\exp(-A/35+1.2)$ MeV, and $\delta E=10.0$ MeV, which is
simple and fulfills this requirement.

In order to check the extent to which the main properties of the model, particularly at high energies, are impacted by the replacement  of Eq.\ (\ref{eq:rhoSMM}) by
Eq.\ (\ref{eq:rhom}), the top panel of Fig.\ \ref{fig:aveobs} displays the primary and final charge multiplicities as a function of the excitation energy for the breakup of the $^{40}$Ca nucleus, whereas the bottom panel exhibits the corresponding caloric curve.
The breakup temperature $T$ is calculated through:

\begin{equation}
\frac{1}{T}=\frac{\partial\ln(\Omega_{A_0,Z_0,Q})}{\partial (Q\Delta_Q)}\approx \frac{\ln(\Omega_{A_0,Z_0,Q})-\ln(\Omega_{A_0,Z_0,Q-1})}{\Delta_Q}\;.
\label{eq:disc}
\end{equation}

\noindent
The similarity of the results shown in  Fig.\ \ref{fig:aveobs} strongly suggests that this improved state density can be safely used and so it is done from this point on.

\end{section}

\begin{figure}[htb]
\includegraphics[width=8.5cm,angle=0]{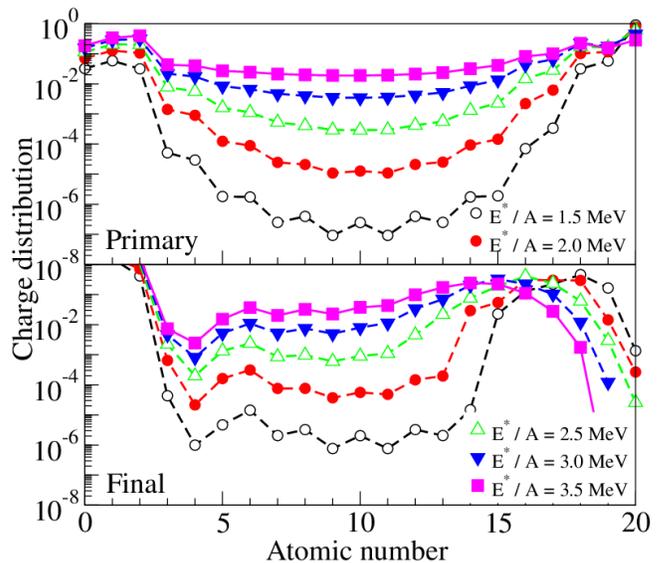}
\caption{\label{fig:cdwp} (Color online) Same as \ref{fig:smmpw} but the state density is replaced by Eq.\ (\ref{eq:rhom}).
For details see the text.}
\end{figure}

\begin{section}{Results}
\label{sect:results}
We now investigate the role played by the improved state density given by Eq.\ (\ref{eq:rhom}) in the staggering properties of the charge distribution.
In this way, the primary and final charge distributions for the fragmentation of the $^{40}$Ca nucleus, at a few excitation energies, are displayed in Fig.\ \ref{fig:cdwp}.
In contrast with the previous results shown in Fig.\ \ref{fig:smmpw}, odd-even effects in the primary distribution quickly disappear as the excitation energy increases, being barely noticeable at $E^*/A\approx 2.5$ MeV.
On the other hand, these effects are significantly enhanced in the final yields, as already suggested in former studies \cite{StaggeringRicciardi2005,Staggering2013,StaggeringDagostino2011}.
They also tend to be smoothed out as $Z$ increases, in agreement with the findings of Ref.\ \cite{StaggeringCasini2013}.

In order to examine the dependence of the staggering on the source's size, we also consider the breakup of the $^{80}$Zr nucleus.
The corresponding primary and final charge distributions are exhibited in Fig.\ \ref{fig:cd80Zrwp}.
The smoothing of the charge distribution is much more accentuated in this case than in the fragmentation of the $^{40}$Ca nucleus.
The magnitude of the staggering is very small at $E^*/A=1.5$ MeV and disappears almost completely at slightly higher excitation energies.
The effects are somewhat more important at lower excitation energies, but the yields are extremely small.

\begin{figure}[tbh]
\includegraphics[width=8.5cm,angle=0]{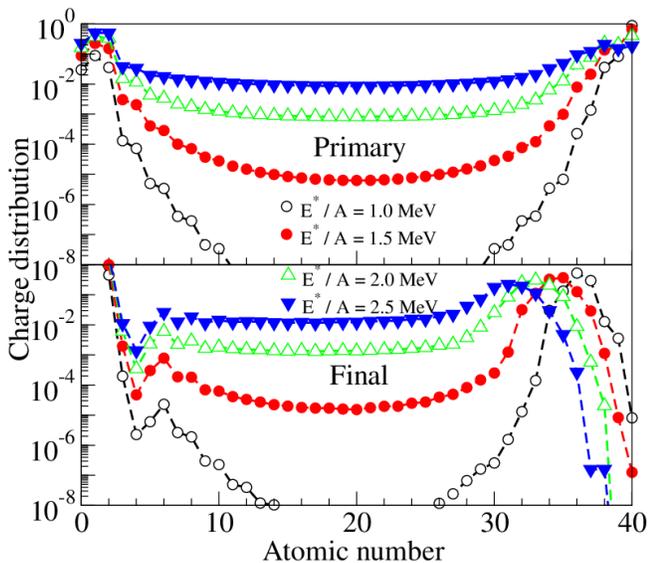}
\caption{\label{fig:cd80Zrwp} (Color online) Same as Fig.\ \ref{fig:cdwp} for the $^{80}$Zr nucleus.
For details see the text.}
\end{figure}

\begin{figure}[tbh]
\includegraphics[width=8.5cm,angle=0]{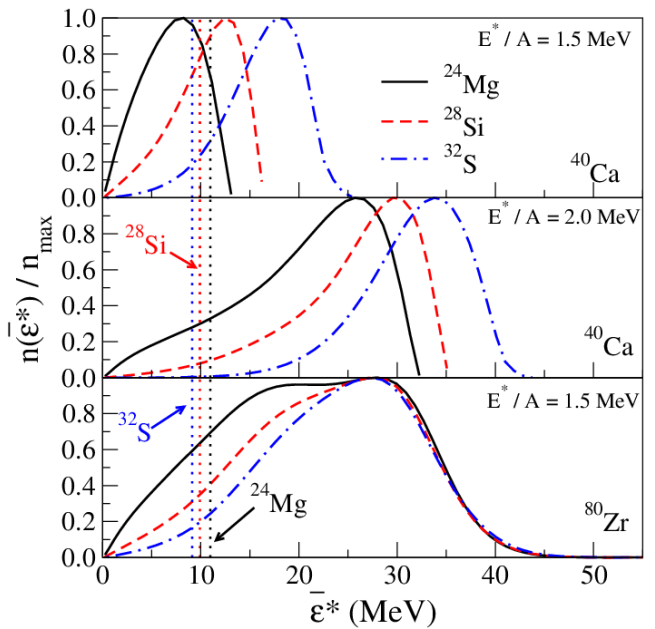}
\caption{\label{fig:avee} (Color online) Average excitation energy of the primary fragments produced in the breakup of the $^{40}$Ca and $^{80}$Zr nuclei.
The vertical dotted lines represent $E_x$ for $^{24}$Mg, $^{28}$Si, and $^{32}$S, as indicated by the arrows and labels.
For details see the text.}
\end{figure}

The dependence of the odd-even staggering on the excitation energy, as well as on the fragment's and system's size, can be understood by examining the excitation energy distribution of the primary fragments.
This  is calculated through:

\begin{equation}
\overline{\epsilon}^*=\frac{\gamma_a}{\omega_{a,z,q}}\int_0^{\epsilon_{a,z,q}}\,dK\;(\epsilon_{a,z,q}-K)\sqrt{K}\rho(\epsilon_{a,z,q}-K)\;.
\label{eq:eex}
\end{equation}

\noindent
The results are shown in Fig.\ \ref{fig:avee}, for the $^{24}$Mg, $^{28}$Si, and $^{32}$S fragments, produced in the breakup of the $^{40}$Ca and $^{80}$Zr nuclei.
The top panel of this figure displays the results corresponding to the $^{40}$Ca  source  at $E^*/A=1.5$~MeV.
For each of the considered fragments, the vertical dotted lines represent $E_x$, which is the energy value below which collective states play a relevant role.
One sees that the fraction of fragments produced with excitation energy higher than $E_x$ quickly increases with the fragment's size.
Therefore, pairing effects should become progressively less important as the fragment's size increases.

The middle panel shows that the excitation energy of the source also plays a very important role.
Indeed, $E^*/A$ increases only by 0.5 MeV from the top to the middle panel but the fraction of fragments with excitation energy below $E_x$ decreases substantially.
This should significantly weaken the odd-even effects on the charge distribution as is indeed noticed in Fig.\ \ref{fig:cdwp}.

In the bottom panel of Fig.\ \ref{fig:avee} is shown the average excitation energy for the same fragments, but for a $^{80}$Zr nucleus as a source.
The excitation energy, $E^*/A=1.5$ MeV, is the same as in the case of the top panel, for the $^{40}$Ca source.
The much larger amount of the avaliable excitation energy causes the energy distribution to become broader and its peak to move to higher excitation energies.
Once more, the population with excitation energy below $E_x$ is significantly reduced, leading to smoother charge distributions.
One should note that the breakup temperature is $T=3.89$ MeV, which is very close to $T=3.74$ MeV obtained with the $^{40}$Ca source at the same excitation energy.
These results reveal that, in this context, the total excitation energy is the relevant quantity, instead of the excitation energy per nucleon.

Thus, in the framework of the SMM, the excitation energy, fragment's size and source's size dependence are explained by the migration of the population in low-lying to high-lying states.
Therefore, the study of the odd-even staggering may help to obtain information on the nuclear state density.

\begin{figure}[tbh]
\includegraphics[width=8.5cm,angle=0]{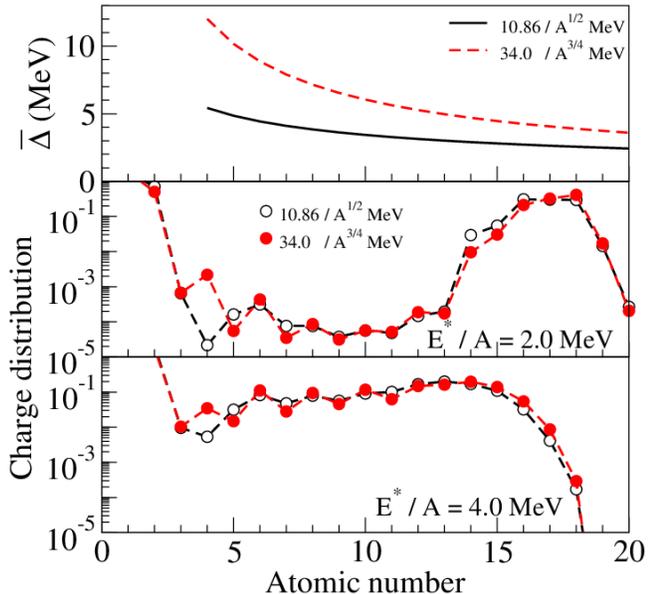}
\caption{\label{fig:esym} (Color online) Top panel: comparison between different pairing energies used in this work. Middle and bottom panels: Comparison between the final charge distributions for the fragmentation of $^{40}$Ca  obtained with the different symmetry energy parameterizations.
For details see the text.}
\end{figure}

Finally, we examine the extent to which it is possible to distinguish between different parameterizations of the pairing energy from this analysis.
Besides that adopted in SMM, which amplitude is $\overline\Delta\equiv 11.86/A^{1/2}$ MeV, we also carried out the calculations with the pairing term used in \cite{rs80}, in which case the amplitude reads $\overline\Delta\equiv 34.0/A^{3/4}$.
The comparison between these two terms is shown in the upper panel of Fig.\ \ref{fig:esym}, from which one sees that these  prescriptions lead to important differences between the pairing energies.
Notwithstanding this, the influence of this change of parameterization on the primary charge distribution is very small, so they are not shown in this figure.
The differences are amplified in the deexcitation process, as is illustrated in the middle and bottom panels of Fig.\ \ref{fig:esym}, which show the final charge distribution for the fragmentation of the $^{40}$Ca nucleus at $E^*/A=2.0$ MeV and $E^*/A=4.0$ MeV.
The larger pairing energy clearly leads to more important odd-even effects and, therefore, the analyses made in Refs.\ \cite{StaggeringRicciardi2005,StaggeringDagostino2011,StaggeringCasini2012,Staggering2013,StaggeringCasini2013} may be helpful in finding the best parameterization for the pairing term, but this requires a careful treatment of the deexcitation of the primary fragments in order to minimize ambiguities.

\end{section}

\begin{section}{Concluding remarks}
\label{sect:conclusions}
By modifying the density of states employed in the SMM, in order to take the pairing energy into account, we have studied the odd-even staggering in the charge distribution of fragments produced in the breakup of excited nuclear systems.

In agreement with previous results \cite{StaggeringRicciardi2005,Staggering2013,StaggeringDagostino2011}, we find that this staggering is strongly influenced by the deexcitation of the primary hot fragments, so that it can be useful in tuning the treatments used to describe this stage of the process.

The smoothing of the charge distribution of primary fragments is explained in the framework of our model by the increasing of the population in states of energies for which the excitation of single particle states becomes dominant in comparison to collective modes, being thus well described by a Fermi gas.
Since the density of states is one of the ingredients of the calculations, our results suggest that the odd-even effects observed in the charge distribution may also be used to constrain this quantity.

We also found that a careful comparison with the experimental results may help to distinguish between different parameterizations of the pairing energy.

In conclusion, the sensitivity of this odd-even staggering to the key ingredients of the statistical calculations should  be very useful in improving these models, constraining some of their parameters and assumptions.

\end{section}

\begin{acknowledgments}
We would like to acknowledge CNPq, FAPERJ BBP grant, FAPESP and the joint PRONEX initiatives of CNPq/FAPERJ under
Contract No.\ 26-111.443/2010, for partial financial support.
This work was supported in part by the National Science Foundation under Grant No. PHY-1102511.
We also thank the
Programa de Desarrollo de las Ciencias B\'asicas (PEDECIBA) and the
Agencia Nacional de Investigaci\'on e Innovaci\'on (ANII) for partial financial support.
\end{acknowledgments}

\bibliography{manuscript}

\begin{thebibliography}{35}%
\makeatletter
\providecommand \@ifxundefined [1]{%
 \@ifx{#1\undefined}
}%
\providecommand \@ifnum [1]{%
 \ifnum #1\expandafter \@firstoftwo
 \else \expandafter \@secondoftwo
 \fi
}%
\providecommand \@ifx [1]{%
 \ifx #1\expandafter \@firstoftwo
 \else \expandafter \@secondoftwo
 \fi
}%
\providecommand \natexlab [1]{#1}%
\providecommand \enquote  [1]{``#1''}%
\providecommand \bibnamefont  [1]{#1}%
\providecommand \bibfnamefont [1]{#1}%
\providecommand \citenamefont [1]{#1}%
\providecommand \href@noop [0]{\@secondoftwo}%
\providecommand \href [0]{\begingroup \@sanitize@url \@href}%
\providecommand \@href[1]{\@@startlink{#1}\@@href}%
\providecommand \@@href[1]{\endgroup#1\@@endlink}%
\providecommand \@sanitize@url [0]{\catcode `\\12\catcode `\$12\catcode
  `\&12\catcode `\#12\catcode `\^12\catcode `\_12\catcode `\%12\relax}%
\providecommand \@@startlink[1]{}%
\providecommand \@@endlink[0]{}%
\providecommand \url  [0]{\begingroup\@sanitize@url \@url }%
\providecommand \@url [1]{\endgroup\@href {#1}{\urlprefix }}%
\providecommand \urlprefix  [0]{URL }%
\providecommand \Eprint [0]{\href }%
\providecommand \doibase [0]{http://dx.doi.org/}%
\providecommand \selectlanguage [0]{\@gobble}%
\providecommand \bibinfo  [0]{\@secondoftwo}%
\providecommand \bibfield  [0]{\@secondoftwo}%
\providecommand \translation [1]{[#1]}%
\providecommand \BibitemOpen [0]{}%
\providecommand \bibitemStop [0]{}%
\providecommand \bibitemNoStop [0]{.\EOS\space}%
\providecommand \EOS [0]{\spacefactor3000\relax}%
\providecommand \BibitemShut  [1]{\csname bibitem#1\endcsname}%
\let\auto@bib@innerbib\@empty
\bibitem [{\citenamefont {Moretto}\ and\ \citenamefont
  {Wozniak}(1993)}]{Moretto1993}%
  \BibitemOpen
  \bibfield  {author} {\bibinfo {author} {\bibfnamefont {L.~G.}\ \bibnamefont
  {Moretto}}\ and\ \bibinfo {author} {\bibfnamefont {G.~J.}\ \bibnamefont
  {Wozniak}},\ }\href@noop {} {\bibfield  {journal} {\bibinfo  {journal} {Ann.
  Rev. Nucl. Part. Sci.}\ }\textbf {\bibinfo {volume} {43}},\ \bibinfo {pages}
  {379} (\bibinfo {year} {1993})}\BibitemShut {NoStop}%
\bibitem [{\citenamefont {Bondorf}\ \emph {et~al.}(1995)\citenamefont
  {Bondorf}, \citenamefont {Botvina}, \citenamefont {Iljinov}, \citenamefont
  {Mihustin},\ and\ \citenamefont {Sneppen}}]{Bondorf1995}%
  \BibitemOpen
  \bibfield  {author} {\bibinfo {author} {\bibfnamefont {J.~P.}\ \bibnamefont
  {Bondorf}}, \bibinfo {author} {\bibfnamefont {A.~S.}\ \bibnamefont
  {Botvina}}, \bibinfo {author} {\bibfnamefont {A.~S.}\ \bibnamefont
  {Iljinov}}, \bibinfo {author} {\bibfnamefont {I.~N.}\ \bibnamefont
  {Mihustin}}, \ and\ \bibinfo {author} {\bibfnamefont {K.}~\bibnamefont
  {Sneppen}},\ }\href@noop {} {\bibfield  {journal} {\bibinfo  {journal} {Phys.
  Rep.}\ }\textbf {\bibinfo {volume} {257}},\ \bibinfo {pages} {133} (\bibinfo
  {year} {1995})}\BibitemShut {NoStop}%
\bibitem [{\citenamefont {Pochodzalla}(1997)}]{PochodzallaReview1997}%
  \BibitemOpen
  \bibfield  {author} {\bibinfo {author} {\bibfnamefont {J.}~\bibnamefont
  {Pochodzalla}},\ }\href {\doibase
  http://dx.doi.org/10.1016/S0146-6410(97)00048-3} {\bibfield  {journal}
  {\bibinfo  {journal} {Progress in Particle and Nuclear Physics}\ }\textbf
  {\bibinfo {volume} {39}},\ \bibinfo {pages} {443 } (\bibinfo {year}
  {1997})}\BibitemShut {NoStop}%
\bibitem [{\citenamefont {Tan}\ \emph {et~al.}(2001)\citenamefont {Tan},
  \citenamefont {Li}, \citenamefont {Donangelo}, \citenamefont {Gelbke},
  \citenamefont {vanGoethem}, \citenamefont {Liu}, \citenamefont {Lynch},
  \citenamefont {Souza}, \citenamefont {Tsang}, \citenamefont {Verde},
  \citenamefont {Wagner},\ and\ \citenamefont {Xu}}]{TanIsoEOS}%
  \BibitemOpen
  \bibfield  {author} {\bibinfo {author} {\bibfnamefont {W.~P.}\ \bibnamefont
  {Tan}}, \bibinfo {author} {\bibfnamefont {B.-A.}\ \bibnamefont {Li}},
  \bibinfo {author} {\bibfnamefont {R.}~\bibnamefont {Donangelo}}, \bibinfo
  {author} {\bibfnamefont {C.~K.}\ \bibnamefont {Gelbke}}, \bibinfo {author}
  {\bibfnamefont {M.-J.}\ \bibnamefont {vanGoethem}}, \bibinfo {author}
  {\bibfnamefont {X.~D.}\ \bibnamefont {Liu}}, \bibinfo {author} {\bibfnamefont
  {W.~G.}\ \bibnamefont {Lynch}}, \bibinfo {author} {\bibfnamefont
  {S.}~\bibnamefont {Souza}}, \bibinfo {author} {\bibfnamefont {M.~B.}\
  \bibnamefont {Tsang}}, \bibinfo {author} {\bibfnamefont {G.}~\bibnamefont
  {Verde}}, \bibinfo {author} {\bibfnamefont {A.}~\bibnamefont {Wagner}}, \
  and\ \bibinfo {author} {\bibfnamefont {H.~S.}\ \bibnamefont {Xu}},\ }\href
  {\doibase 10.1103/PhysRevC.64.051901} {\bibfield  {journal} {\bibinfo
  {journal} {Phys. Rev. C}\ }\textbf {\bibinfo {volume} {64}},\ \bibinfo
  {pages} {051901} (\bibinfo {year} {2001})}\BibitemShut {NoStop}%
\bibitem [{\citenamefont {Ogilvie}\ \emph {et~al.}(1991)\citenamefont
  {Ogilvie}, \citenamefont {Adloff}, \citenamefont {Begemann-Blaich},
  \citenamefont {Bouissou}, \citenamefont {Hubele}, \citenamefont {Imme},
  \citenamefont {Iori}, \citenamefont {Kreutz}, \citenamefont {Kunde},
  \citenamefont {Leray}, \citenamefont {Lindenstruth}, \citenamefont {Liu},
  \citenamefont {Lynen}, \citenamefont {Meijer}, \citenamefont {Milkau},
  \citenamefont {M\"uller}, \citenamefont {Ng\^o}, \citenamefont {Pochodzalla},
  \citenamefont {Raciti}, \citenamefont {Rudolf}, \citenamefont {Sann},
  \citenamefont {Sch\"uttauf}, \citenamefont {Seidel}, \citenamefont {Stuttge},
  \citenamefont {Trautmann},\ and\ \citenamefont {Tucholski}}]{riseandfall}%
  \BibitemOpen
  \bibfield  {author} {\bibinfo {author} {\bibfnamefont {C.~A.}\ \bibnamefont
  {Ogilvie}}, \bibinfo {author} {\bibfnamefont {J.~C.}\ \bibnamefont {Adloff}},
  \bibinfo {author} {\bibfnamefont {M.}~\bibnamefont {Begemann-Blaich}},
  \bibinfo {author} {\bibfnamefont {P.}~\bibnamefont {Bouissou}}, \bibinfo
  {author} {\bibfnamefont {J.}~\bibnamefont {Hubele}}, \bibinfo {author}
  {\bibfnamefont {G.}~\bibnamefont {Imme}}, \bibinfo {author} {\bibfnamefont
  {I.}~\bibnamefont {Iori}}, \bibinfo {author} {\bibfnamefont {P.}~\bibnamefont
  {Kreutz}}, \bibinfo {author} {\bibfnamefont {G.~J.}\ \bibnamefont {Kunde}},
  \bibinfo {author} {\bibfnamefont {S.}~\bibnamefont {Leray}}, \bibinfo
  {author} {\bibfnamefont {V.}~\bibnamefont {Lindenstruth}}, \bibinfo {author}
  {\bibfnamefont {Z.}~\bibnamefont {Liu}}, \bibinfo {author} {\bibfnamefont
  {U.}~\bibnamefont {Lynen}}, \bibinfo {author} {\bibfnamefont {R.~J.}\
  \bibnamefont {Meijer}}, \bibinfo {author} {\bibfnamefont {U.}~\bibnamefont
  {Milkau}}, \bibinfo {author} {\bibfnamefont {W.~F.~J.}\ \bibnamefont
  {M\"uller}}, \bibinfo {author} {\bibfnamefont {C.}~\bibnamefont {Ng\^o}},
  \bibinfo {author} {\bibfnamefont {J.}~\bibnamefont {Pochodzalla}}, \bibinfo
  {author} {\bibfnamefont {G.}~\bibnamefont {Raciti}}, \bibinfo {author}
  {\bibfnamefont {G.}~\bibnamefont {Rudolf}}, \bibinfo {author} {\bibfnamefont
  {H.}~\bibnamefont {Sann}}, \bibinfo {author} {\bibfnamefont {A.}~\bibnamefont
  {Sch\"uttauf}}, \bibinfo {author} {\bibfnamefont {W.}~\bibnamefont {Seidel}},
  \bibinfo {author} {\bibfnamefont {L.}~\bibnamefont {Stuttge}}, \bibinfo
  {author} {\bibfnamefont {W.}~\bibnamefont {Trautmann}}, \ and\ \bibinfo
  {author} {\bibfnamefont {A.}~\bibnamefont {Tucholski}},\ }\href {\doibase
  10.1103/PhysRevLett.67.1214} {\bibfield  {journal} {\bibinfo  {journal}
  {Phys. Rev. Lett.}\ }\textbf {\bibinfo {volume} {67}},\ \bibinfo {pages}
  {1214} (\bibinfo {year} {1991})}\BibitemShut {NoStop}%
\bibitem [{\citenamefont {Peaslee}\ \emph {et~al.}(1994)\citenamefont
  {Peaslee}, \citenamefont {Tsang}, \citenamefont {Schwarz}, \citenamefont
  {Huang}, \citenamefont {Huang}, \citenamefont {Hsi}, \citenamefont
  {Williams}, \citenamefont {Bauer}, \citenamefont {Bowman}, \citenamefont
  {Chartier}, \citenamefont {Dinius}, \citenamefont {Gelbke}, \citenamefont
  {Glasmacher}, \citenamefont {Handzy}, \citenamefont {Lisa}, \citenamefont
  {Lynch}, \citenamefont {Mader}, \citenamefont {Phair}, \citenamefont
  {Lemaire}, \citenamefont {Souza}, \citenamefont {Van~Buren}, \citenamefont
  {Charity}, \citenamefont {Sobotka}, \citenamefont {Kunde}, \citenamefont
  {Lynen}, \citenamefont {Pochodzalla}, \citenamefont {Sann}, \citenamefont
  {Trautmann}, \citenamefont {Fox}, \citenamefont {de~Souza}, \citenamefont
  {Peilert}, \citenamefont {Friedman},\ and\ \citenamefont
  {Carlin}}]{Peaslee1994}%
  \BibitemOpen
  \bibfield  {author} {\bibinfo {author} {\bibfnamefont {G.~F.}\ \bibnamefont
  {Peaslee}}, \bibinfo {author} {\bibfnamefont {M.~B.}\ \bibnamefont {Tsang}},
  \bibinfo {author} {\bibfnamefont {C.}~\bibnamefont {Schwarz}}, \bibinfo
  {author} {\bibfnamefont {M.~J.}\ \bibnamefont {Huang}}, \bibinfo {author}
  {\bibfnamefont {W.~S.}\ \bibnamefont {Huang}}, \bibinfo {author}
  {\bibfnamefont {W.~C.}\ \bibnamefont {Hsi}}, \bibinfo {author} {\bibfnamefont
  {C.}~\bibnamefont {Williams}}, \bibinfo {author} {\bibfnamefont
  {W.}~\bibnamefont {Bauer}}, \bibinfo {author} {\bibfnamefont {D.~R.}\
  \bibnamefont {Bowman}}, \bibinfo {author} {\bibfnamefont {M.}~\bibnamefont
  {Chartier}}, \bibinfo {author} {\bibfnamefont {J.}~\bibnamefont {Dinius}},
  \bibinfo {author} {\bibfnamefont {C.~K.}\ \bibnamefont {Gelbke}}, \bibinfo
  {author} {\bibfnamefont {T.}~\bibnamefont {Glasmacher}}, \bibinfo {author}
  {\bibfnamefont {D.~O.}\ \bibnamefont {Handzy}}, \bibinfo {author}
  {\bibfnamefont {M.~A.}\ \bibnamefont {Lisa}}, \bibinfo {author}
  {\bibfnamefont {W.~G.}\ \bibnamefont {Lynch}}, \bibinfo {author}
  {\bibfnamefont {C.~M.}\ \bibnamefont {Mader}}, \bibinfo {author}
  {\bibfnamefont {L.}~\bibnamefont {Phair}}, \bibinfo {author} {\bibfnamefont
  {M.-C.}\ \bibnamefont {Lemaire}}, \bibinfo {author} {\bibfnamefont {S.~R.}\
  \bibnamefont {Souza}}, \bibinfo {author} {\bibfnamefont {G.}~\bibnamefont
  {Van~Buren}}, \bibinfo {author} {\bibfnamefont {R.~J.}\ \bibnamefont
  {Charity}}, \bibinfo {author} {\bibfnamefont {L.~G.}\ \bibnamefont
  {Sobotka}}, \bibinfo {author} {\bibfnamefont {G.~J.}\ \bibnamefont {Kunde}},
  \bibinfo {author} {\bibfnamefont {U.}~\bibnamefont {Lynen}}, \bibinfo
  {author} {\bibfnamefont {J.}~\bibnamefont {Pochodzalla}}, \bibinfo {author}
  {\bibfnamefont {H.}~\bibnamefont {Sann}}, \bibinfo {author} {\bibfnamefont
  {W.}~\bibnamefont {Trautmann}}, \bibinfo {author} {\bibfnamefont
  {D.}~\bibnamefont {Fox}}, \bibinfo {author} {\bibfnamefont {R.~T.}\
  \bibnamefont {de~Souza}}, \bibinfo {author} {\bibfnamefont {G.}~\bibnamefont
  {Peilert}}, \bibinfo {author} {\bibfnamefont {W.~A.}\ \bibnamefont
  {Friedman}}, \ and\ \bibinfo {author} {\bibfnamefont {N.}~\bibnamefont
  {Carlin}},\ }\href {\doibase 10.1103/PhysRevC.49.R2271} {\bibfield  {journal}
  {\bibinfo  {journal} {Phys. Rev. C}\ }\textbf {\bibinfo {volume} {49}},\
  \bibinfo {pages} {R2271} (\bibinfo {year} {1994})}\BibitemShut {NoStop}%
\bibitem [{\citenamefont {{Das~Gupta}}\ \emph {et~al.}(2001)\citenamefont
  {{Das~Gupta}}, \citenamefont {Mekjian},\ and\ \citenamefont
  {Tsang}}]{reviewSubal2001}%
  \BibitemOpen
  \bibfield  {author} {\bibinfo {author} {\bibfnamefont {S.}~\bibnamefont
  {{Das~Gupta}}}, \bibinfo {author} {\bibfnamefont {A.~Z.}\ \bibnamefont
  {Mekjian}}, \ and\ \bibinfo {author} {\bibfnamefont {M.~B.}\ \bibnamefont
  {Tsang}},\ }\href@noop {} {\bibfield  {journal} {\bibinfo  {journal}
  {Advances in Nuclear Physics}\ }\textbf {\bibinfo {volume} {26}},\ \bibinfo
  {pages} {89} (\bibinfo {year} {2001})}\BibitemShut {NoStop}%
\bibitem [{\citenamefont {Das}\ \emph {et~al.}(2005)\citenamefont {Das},
  \citenamefont {{Das~Gupta}}, \citenamefont {Lynch}, \citenamefont {Mekjian},\
  and\ \citenamefont {Tsang}}]{BettyPhysRep2005}%
  \BibitemOpen
  \bibfield  {author} {\bibinfo {author} {\bibfnamefont {C.~B.}\ \bibnamefont
  {Das}}, \bibinfo {author} {\bibfnamefont {S.}~\bibnamefont {{Das~Gupta}}},
  \bibinfo {author} {\bibfnamefont {W.~G.}\ \bibnamefont {Lynch}}, \bibinfo
  {author} {\bibfnamefont {A.~Z.}\ \bibnamefont {Mekjian}}, \ and\ \bibinfo
  {author} {\bibfnamefont {M.~B.}\ \bibnamefont {Tsang}},\ }\href@noop {}
  {\bibfield  {journal} {\bibinfo  {journal} {Phys. Rep.}\ }\textbf {\bibinfo
  {volume} {406}},\ \bibinfo {pages} {1} (\bibinfo {year} {2005})}\BibitemShut
  {NoStop}%
\bibitem [{\citenamefont {Xu}\ \emph {et~al.}(2000)\citenamefont {Xu},
  \citenamefont {Tsang}, \citenamefont {Liu}, \citenamefont {Liu},
  \citenamefont {Lynch}, \citenamefont {Tan}, \citenamefont {{Vander~Molen}},
  \citenamefont {Verde}, \citenamefont {Wagner}, \citenamefont {Xi},
  \citenamefont {Gelbke}, \citenamefont {Beaulieu}, \citenamefont {Davin},
  \citenamefont {Larochelle}, \citenamefont {Lefort}, \citenamefont
  {{de~Souza}}, \citenamefont {Yanez}, \citenamefont {Viola}, \citenamefont
  {Charity},\ and\ \citenamefont {Sobotka}}]{isoscaling1}%
  \BibitemOpen
  \bibfield  {author} {\bibinfo {author} {\bibfnamefont {H.~S.}\ \bibnamefont
  {Xu}}, \bibinfo {author} {\bibfnamefont {M.~B.}\ \bibnamefont {Tsang}},
  \bibinfo {author} {\bibfnamefont {T.~X.}\ \bibnamefont {Liu}}, \bibinfo
  {author} {\bibfnamefont {X.~D.}\ \bibnamefont {Liu}}, \bibinfo {author}
  {\bibfnamefont {W.~G.}\ \bibnamefont {Lynch}}, \bibinfo {author}
  {\bibfnamefont {W.~P.}\ \bibnamefont {Tan}}, \bibinfo {author} {\bibfnamefont
  {A.}~\bibnamefont {{Vander~Molen}}}, \bibinfo {author} {\bibfnamefont
  {G.}~\bibnamefont {Verde}}, \bibinfo {author} {\bibfnamefont
  {A.}~\bibnamefont {Wagner}}, \bibinfo {author} {\bibfnamefont {H.~F.}\
  \bibnamefont {Xi}}, \bibinfo {author} {\bibfnamefont {C.~K.}\ \bibnamefont
  {Gelbke}}, \bibinfo {author} {\bibfnamefont {L.}~\bibnamefont {Beaulieu}},
  \bibinfo {author} {\bibfnamefont {B.}~\bibnamefont {Davin}}, \bibinfo
  {author} {\bibfnamefont {Y.}~\bibnamefont {Larochelle}}, \bibinfo {author}
  {\bibfnamefont {T.}~\bibnamefont {Lefort}}, \bibinfo {author} {\bibfnamefont
  {R.~T.}\ \bibnamefont {{de~Souza}}}, \bibinfo {author} {\bibfnamefont
  {R.}~\bibnamefont {Yanez}}, \bibinfo {author} {\bibfnamefont {V.~E.}\
  \bibnamefont {Viola}}, \bibinfo {author} {\bibfnamefont {R.~J.}\ \bibnamefont
  {Charity}}, \ and\ \bibinfo {author} {\bibfnamefont {L.~G.}\ \bibnamefont
  {Sobotka}},\ }\href@noop {} {\bibfield  {journal} {\bibinfo  {journal} {Phys.
  Rev. Lett.}\ }\textbf {\bibinfo {volume} {85}},\ \bibinfo {pages} {716}
  (\bibinfo {year} {2000})}\BibitemShut {NoStop}%
\bibitem [{\citenamefont {Tsang}\ \emph
  {et~al.}(2001{\natexlab{a}})\citenamefont {Tsang}, \citenamefont {Friedman},
  \citenamefont {Gelbke}, \citenamefont {Lynch}, \citenamefont {Verde},\ and\
  \citenamefont {Xu}}]{isoscaling2}%
  \BibitemOpen
  \bibfield  {author} {\bibinfo {author} {\bibfnamefont {M.~B.}\ \bibnamefont
  {Tsang}}, \bibinfo {author} {\bibfnamefont {W.~A.}\ \bibnamefont {Friedman}},
  \bibinfo {author} {\bibfnamefont {C.~K.}\ \bibnamefont {Gelbke}}, \bibinfo
  {author} {\bibfnamefont {W.~G.}\ \bibnamefont {Lynch}}, \bibinfo {author}
  {\bibfnamefont {G.}~\bibnamefont {Verde}}, \ and\ \bibinfo {author}
  {\bibfnamefont {H.~S.}\ \bibnamefont {Xu}},\ }\href@noop {} {\bibfield
  {journal} {\bibinfo  {journal} {Phys. Rev. Lett.}\ }\textbf {\bibinfo
  {volume} {86}},\ \bibinfo {pages} {5023} (\bibinfo {year}
  {2001}{\natexlab{a}})}\BibitemShut {NoStop}%
\bibitem [{\citenamefont {Souza}\ \emph {et~al.}(2004)\citenamefont {Souza},
  \citenamefont {Donangelo}, \citenamefont {Lynch}, \citenamefont {Tan},\ and\
  \citenamefont {Tsang}}]{isocc}%
  \BibitemOpen
  \bibfield  {author} {\bibinfo {author} {\bibfnamefont {S.~R.}\ \bibnamefont
  {Souza}}, \bibinfo {author} {\bibfnamefont {R.}~\bibnamefont {Donangelo}},
  \bibinfo {author} {\bibfnamefont {W.~G.}\ \bibnamefont {Lynch}}, \bibinfo
  {author} {\bibfnamefont {W.~P.}\ \bibnamefont {Tan}}, \ and\ \bibinfo
  {author} {\bibfnamefont {M.~B.}\ \bibnamefont {Tsang}},\ }\href {\doibase
  10.1103/PhysRevC.69.031607} {\bibfield  {journal} {\bibinfo  {journal} {Phys.
  Rev. C}\ }\textbf {\bibinfo {volume} {69}},\ \bibinfo {pages} {031607(R)}
  (\bibinfo {year} {2004})}\BibitemShut {NoStop}%
\bibitem [{\citenamefont {Tsang}\ \emph
  {et~al.}(2001{\natexlab{b}})\citenamefont {Tsang}, \citenamefont {Gelbke},
  \citenamefont {Liu}, \citenamefont {Lynch}, \citenamefont {Tan},
  \citenamefont {Verde}, \citenamefont {Xu}, \citenamefont {Friedman},
  \citenamefont {Donangelo}, \citenamefont {Souza}, \citenamefont {Das},
  \citenamefont {Das~Gupta},\ and\ \citenamefont {Zhabinsky}}]{isoscaling3}%
  \BibitemOpen
  \bibfield  {author} {\bibinfo {author} {\bibfnamefont {M.~B.}\ \bibnamefont
  {Tsang}}, \bibinfo {author} {\bibfnamefont {C.~K.}\ \bibnamefont {Gelbke}},
  \bibinfo {author} {\bibfnamefont {X.~D.}\ \bibnamefont {Liu}}, \bibinfo
  {author} {\bibfnamefont {W.~G.}\ \bibnamefont {Lynch}}, \bibinfo {author}
  {\bibfnamefont {W.~P.}\ \bibnamefont {Tan}}, \bibinfo {author} {\bibfnamefont
  {G.}~\bibnamefont {Verde}}, \bibinfo {author} {\bibfnamefont {H.~S.}\
  \bibnamefont {Xu}}, \bibinfo {author} {\bibfnamefont {W.~A.}\ \bibnamefont
  {Friedman}}, \bibinfo {author} {\bibfnamefont {R.}~\bibnamefont {Donangelo}},
  \bibinfo {author} {\bibfnamefont {S.~R.}\ \bibnamefont {Souza}}, \bibinfo
  {author} {\bibfnamefont {C.~B.}\ \bibnamefont {Das}}, \bibinfo {author}
  {\bibfnamefont {S.}~\bibnamefont {Das~Gupta}}, \ and\ \bibinfo {author}
  {\bibfnamefont {D.}~\bibnamefont {Zhabinsky}},\ }\href {\doibase
  10.1103/PhysRevC.64.054615} {\bibfield  {journal} {\bibinfo  {journal} {Phys.
  Rev. C}\ }\textbf {\bibinfo {volume} {64}},\ \bibinfo {pages} {054615}
  (\bibinfo {year} {2001}{\natexlab{b}})}\BibitemShut {NoStop}%
\bibitem [{\citenamefont {Souza}\ \emph {et~al.}(2008)\citenamefont {Souza},
  \citenamefont {Tsang}, \citenamefont {Donangelo}, \citenamefont {Lynch},\
  and\ \citenamefont {Steiner}}]{isoMassFormula2008}%
  \BibitemOpen
  \bibfield  {author} {\bibinfo {author} {\bibfnamefont {S.~R.}\ \bibnamefont
  {Souza}}, \bibinfo {author} {\bibfnamefont {M.~B.}\ \bibnamefont {Tsang}},
  \bibinfo {author} {\bibfnamefont {R.}~\bibnamefont {Donangelo}}, \bibinfo
  {author} {\bibfnamefont {W.~G.}\ \bibnamefont {Lynch}}, \ and\ \bibinfo
  {author} {\bibfnamefont {A.~W.}\ \bibnamefont {Steiner}},\ }\href {\doibase
  10.1103/PhysRevC.78.014605} {\bibfield  {journal} {\bibinfo  {journal} {Phys.
  Rev. C}\ }\textbf {\bibinfo {volume} {78}},\ \bibinfo {pages} {014605}
  (\bibinfo {year} {2008})}\BibitemShut {NoStop}%
\bibitem [{\citenamefont {Souza}\ \emph
  {et~al.}(2009{\natexlab{a}})\citenamefont {Souza}, \citenamefont {Tsang},
  \citenamefont {Carlson}, \citenamefont {Donangelo}, \citenamefont {Lynch},\
  and\ \citenamefont {Steiner}}]{isoSMMTF}%
  \BibitemOpen
  \bibfield  {author} {\bibinfo {author} {\bibfnamefont {S.~R.}\ \bibnamefont
  {Souza}}, \bibinfo {author} {\bibfnamefont {M.~B.}\ \bibnamefont {Tsang}},
  \bibinfo {author} {\bibfnamefont {B.~V.}\ \bibnamefont {Carlson}}, \bibinfo
  {author} {\bibfnamefont {R.}~\bibnamefont {Donangelo}}, \bibinfo {author}
  {\bibfnamefont {W.~G.}\ \bibnamefont {Lynch}}, \ and\ \bibinfo {author}
  {\bibfnamefont {A.~W.}\ \bibnamefont {Steiner}},\ }\href {\doibase
  10.1103/PhysRevC.80.041602} {\bibfield  {journal} {\bibinfo  {journal} {Phys.
  Rev. C}\ }\textbf {\bibinfo {volume} {80}},\ \bibinfo {pages} {041602(R)}
  (\bibinfo {year} {2009}{\natexlab{a}})}\BibitemShut {NoStop}%
\bibitem [{\citenamefont {Souza}\ \emph
  {et~al.}(2009{\natexlab{b}})\citenamefont {Souza}, \citenamefont {Tsang},
  \citenamefont {Carlson}, \citenamefont {Donangelo}, \citenamefont {Lynch},\
  and\ \citenamefont {Steiner}}]{isotemp}%
  \BibitemOpen
  \bibfield  {author} {\bibinfo {author} {\bibfnamefont {S.~R.}\ \bibnamefont
  {Souza}}, \bibinfo {author} {\bibfnamefont {M.~B.}\ \bibnamefont {Tsang}},
  \bibinfo {author} {\bibfnamefont {B.~V.}\ \bibnamefont {Carlson}}, \bibinfo
  {author} {\bibfnamefont {R.}~\bibnamefont {Donangelo}}, \bibinfo {author}
  {\bibfnamefont {W.~G.}\ \bibnamefont {Lynch}}, \ and\ \bibinfo {author}
  {\bibfnamefont {A.~W.}\ \bibnamefont {Steiner}},\ }\href {\doibase
  10.1103/PhysRevC.80.044606} {\bibfield  {journal} {\bibinfo  {journal} {Phys.
  Rev. C}\ }\textbf {\bibinfo {volume} {80}},\ \bibinfo {pages} {044606}
  (\bibinfo {year} {2009}{\natexlab{b}})}\BibitemShut {NoStop}%
\bibitem [{\citenamefont {Chaudhuri}\ \emph {et~al.}(2009)\citenamefont
  {Chaudhuri}, \citenamefont {Gulminelli},\ and\ \citenamefont {{\protect Das
  Gupta}}}]{isoscalingDasGupta2009}%
  \BibitemOpen
  \bibfield  {author} {\bibinfo {author} {\bibfnamefont {G.}~\bibnamefont
  {Chaudhuri}}, \bibinfo {author} {\bibfnamefont {F.}~\bibnamefont
  {Gulminelli}}, \ and\ \bibinfo {author} {\bibfnamefont {S.}~\bibnamefont
  {{\protect Das Gupta}}},\ }\href {\doibase 10.1103/PhysRevC.80.054606}
  {\bibfield  {journal} {\bibinfo  {journal} {Phys. Rev. C}\ }\textbf {\bibinfo
  {volume} {80}},\ \bibinfo {pages} {054606} (\bibinfo {year}
  {2009})}\BibitemShut {NoStop}%
\bibitem [{\citenamefont {Souza}\ and\ \citenamefont
  {Tsang}(2012)}]{finiteSizeEffects2012}%
  \BibitemOpen
  \bibfield  {author} {\bibinfo {author} {\bibfnamefont {S.~R.}\ \bibnamefont
  {Souza}}\ and\ \bibinfo {author} {\bibfnamefont {M.~B.}\ \bibnamefont
  {Tsang}},\ }\href {\doibase 10.1103/PhysRevC.85.024603} {\bibfield  {journal}
  {\bibinfo  {journal} {Phys. Rev. C}\ }\textbf {\bibinfo {volume} {85}},\
  \bibinfo {pages} {024603} (\bibinfo {year} {2012})}\BibitemShut {NoStop}%
\bibitem [{\citenamefont {Ricciardi}\ \emph {et~al.}(2005)\citenamefont
  {Ricciardi}, \citenamefont {Ignatyuk}, \citenamefont {Keli\'c}, \citenamefont
  {Napolitani}, \citenamefont {Rejmund}, \citenamefont {Schmidt},\ and\
  \citenamefont {Yordanov}}]{StaggeringRicciardi2005}%
  \BibitemOpen
  \bibfield  {author} {\bibinfo {author} {\bibfnamefont {M.}~\bibnamefont
  {Ricciardi}}, \bibinfo {author} {\bibfnamefont {A.}~\bibnamefont {Ignatyuk}},
  \bibinfo {author} {\bibfnamefont {A.}~\bibnamefont {Keli\'c}}, \bibinfo
  {author} {\bibfnamefont {P.}~\bibnamefont {Napolitani}}, \bibinfo {author}
  {\bibfnamefont {F.}~\bibnamefont {Rejmund}}, \bibinfo {author} {\bibfnamefont
  {K.-H.}\ \bibnamefont {Schmidt}}, \ and\ \bibinfo {author} {\bibfnamefont
  {O.}~\bibnamefont {Yordanov}},\ }\href {\doibase
  http://dx.doi.org/10.1016/j.nuclphysa.2004.12.019} {\bibfield  {journal}
  {\bibinfo  {journal} {Nucl. Phys. A}\ }\textbf {\bibinfo {volume} {749}},\
  \bibinfo {pages} {122 } (\bibinfo {year} {2005})}\BibitemShut {NoStop}%
\bibitem [{\citenamefont {D'Agostino}\ \emph {et~al.}(2011)\citenamefont
  {D'Agostino}, \citenamefont {Bruno}, \citenamefont {Gulminelli},
  \citenamefont {Morelli}, \citenamefont {Baiocco}, \citenamefont {Bardelli},
  \citenamefont {Barlini}, \citenamefont {Cannata}, \citenamefont {Casini},
  \citenamefont {Geraci}, \citenamefont {Gramegna}, \citenamefont {Kravchuk},
  \citenamefont {Marchi}, \citenamefont {Moroni}, \citenamefont {Ordine},\ and\
  \citenamefont {Raduta}}]{StaggeringDagostino2011}%
  \BibitemOpen
  \bibfield  {author} {\bibinfo {author} {\bibfnamefont {M.}~\bibnamefont
  {D'Agostino}}, \bibinfo {author} {\bibfnamefont {M.}~\bibnamefont {Bruno}},
  \bibinfo {author} {\bibfnamefont {F.}~\bibnamefont {Gulminelli}}, \bibinfo
  {author} {\bibfnamefont {L.}~\bibnamefont {Morelli}}, \bibinfo {author}
  {\bibfnamefont {G.}~\bibnamefont {Baiocco}}, \bibinfo {author} {\bibfnamefont
  {L.}~\bibnamefont {Bardelli}}, \bibinfo {author} {\bibfnamefont
  {S.}~\bibnamefont {Barlini}}, \bibinfo {author} {\bibfnamefont
  {F.}~\bibnamefont {Cannata}}, \bibinfo {author} {\bibfnamefont
  {G.}~\bibnamefont {Casini}}, \bibinfo {author} {\bibfnamefont
  {E.}~\bibnamefont {Geraci}}, \bibinfo {author} {\bibfnamefont
  {F.}~\bibnamefont {Gramegna}}, \bibinfo {author} {\bibfnamefont
  {V.}~\bibnamefont {Kravchuk}}, \bibinfo {author} {\bibfnamefont
  {T.}~\bibnamefont {Marchi}}, \bibinfo {author} {\bibfnamefont
  {A.}~\bibnamefont {Moroni}}, \bibinfo {author} {\bibfnamefont
  {A.}~\bibnamefont {Ordine}}, \ and\ \bibinfo {author} {\bibfnamefont
  {A.}~\bibnamefont {Raduta}},\ }\href {\doibase
  http://dx.doi.org/10.1016/j.nuclphysa.2011.06.017} {\bibfield  {journal}
  {\bibinfo  {journal} {Nucl. Phys. A}\ }\textbf {\bibinfo {volume} {861}},\
  \bibinfo {pages} {47 } (\bibinfo {year} {2011})}\BibitemShut {NoStop}%
\bibitem [{\citenamefont {Casini}\ \emph {et~al.}(2012)\citenamefont {Casini},
  \citenamefont {Piantelli}, \citenamefont {Maurenzig}, \citenamefont {Olmi},
  \citenamefont {Bardelli}, \citenamefont {Barlini}, \citenamefont {Benelli},
  \citenamefont {Bini}, \citenamefont {Calviani}, \citenamefont {Marini},
  \citenamefont {Mangiarotti}, \citenamefont {Pasquali}, \citenamefont {Poggi},
  \citenamefont {Stefanini}, \citenamefont {Bruno}, \citenamefont {Morelli},
  \citenamefont {Kravchuk}, \citenamefont {Amorini}, \citenamefont {Auditore},
  \citenamefont {Cardella}, \citenamefont {De~Filippo}, \citenamefont
  {Galichet}, \citenamefont {La~Guidara}, \citenamefont {Lanzalone},
  \citenamefont {Lanzan\'o}, \citenamefont {Maiolino}, \citenamefont {Pagano},
  \citenamefont {Papa}, \citenamefont {Pirrone}, \citenamefont {Politi},
  \citenamefont {Pop}, \citenamefont {Porto}, \citenamefont {Rizzo},
  \citenamefont {Russotto}, \citenamefont {Santonocito}, \citenamefont
  {Trifir\'o},\ and\ \citenamefont {Trimarchi}}]{StaggeringCasini2012}%
  \BibitemOpen
  \bibfield  {author} {\bibinfo {author} {\bibfnamefont {G.}~\bibnamefont
  {Casini}}, \bibinfo {author} {\bibfnamefont {S.}~\bibnamefont {Piantelli}},
  \bibinfo {author} {\bibfnamefont {P.~R.}\ \bibnamefont {Maurenzig}}, \bibinfo
  {author} {\bibfnamefont {A.}~\bibnamefont {Olmi}}, \bibinfo {author}
  {\bibfnamefont {L.}~\bibnamefont {Bardelli}}, \bibinfo {author}
  {\bibfnamefont {S.}~\bibnamefont {Barlini}}, \bibinfo {author} {\bibfnamefont
  {M.}~\bibnamefont {Benelli}}, \bibinfo {author} {\bibfnamefont
  {M.}~\bibnamefont {Bini}}, \bibinfo {author} {\bibfnamefont {M.}~\bibnamefont
  {Calviani}}, \bibinfo {author} {\bibfnamefont {P.}~\bibnamefont {Marini}},
  \bibinfo {author} {\bibfnamefont {A.}~\bibnamefont {Mangiarotti}}, \bibinfo
  {author} {\bibfnamefont {G.}~\bibnamefont {Pasquali}}, \bibinfo {author}
  {\bibfnamefont {G.}~\bibnamefont {Poggi}}, \bibinfo {author} {\bibfnamefont
  {A.~A.}\ \bibnamefont {Stefanini}}, \bibinfo {author} {\bibfnamefont
  {M.}~\bibnamefont {Bruno}}, \bibinfo {author} {\bibfnamefont
  {L.}~\bibnamefont {Morelli}}, \bibinfo {author} {\bibfnamefont {V.~L.}\
  \bibnamefont {Kravchuk}}, \bibinfo {author} {\bibfnamefont {F.}~\bibnamefont
  {Amorini}}, \bibinfo {author} {\bibfnamefont {L.}~\bibnamefont {Auditore}},
  \bibinfo {author} {\bibfnamefont {G.}~\bibnamefont {Cardella}}, \bibinfo
  {author} {\bibfnamefont {E.}~\bibnamefont {De~Filippo}}, \bibinfo {author}
  {\bibfnamefont {E.}~\bibnamefont {Galichet}}, \bibinfo {author}
  {\bibfnamefont {E.}~\bibnamefont {La~Guidara}}, \bibinfo {author}
  {\bibfnamefont {G.}~\bibnamefont {Lanzalone}}, \bibinfo {author}
  {\bibfnamefont {G.}~\bibnamefont {Lanzan\'o}}, \bibinfo {author}
  {\bibfnamefont {C.}~\bibnamefont {Maiolino}}, \bibinfo {author}
  {\bibfnamefont {A.}~\bibnamefont {Pagano}}, \bibinfo {author} {\bibfnamefont
  {M.}~\bibnamefont {Papa}}, \bibinfo {author} {\bibfnamefont {S.}~\bibnamefont
  {Pirrone}}, \bibinfo {author} {\bibfnamefont {G.}~\bibnamefont {Politi}},
  \bibinfo {author} {\bibfnamefont {A.}~\bibnamefont {Pop}}, \bibinfo {author}
  {\bibfnamefont {F.}~\bibnamefont {Porto}}, \bibinfo {author} {\bibfnamefont
  {F.}~\bibnamefont {Rizzo}}, \bibinfo {author} {\bibfnamefont
  {P.}~\bibnamefont {Russotto}}, \bibinfo {author} {\bibfnamefont
  {D.}~\bibnamefont {Santonocito}}, \bibinfo {author} {\bibfnamefont
  {A.}~\bibnamefont {Trifir\'o}}, \ and\ \bibinfo {author} {\bibfnamefont
  {M.}~\bibnamefont {Trimarchi}},\ }\href {\doibase 10.1103/PhysRevC.86.011602}
  {\bibfield  {journal} {\bibinfo  {journal} {Phys. Rev. C}\ }\textbf {\bibinfo
  {volume} {86}},\ \bibinfo {pages} {011602} (\bibinfo {year}
  {2012})}\BibitemShut {NoStop}%
\bibitem [{\citenamefont {Winkelbauer}\ \emph {et~al.}(2013)\citenamefont
  {Winkelbauer}, \citenamefont {Souza},\ and\ \citenamefont
  {Tsang}}]{Staggering2013}%
  \BibitemOpen
  \bibfield  {author} {\bibinfo {author} {\bibfnamefont {J.~R.}\ \bibnamefont
  {Winkelbauer}}, \bibinfo {author} {\bibfnamefont {S.~R.}\ \bibnamefont
  {Souza}}, \ and\ \bibinfo {author} {\bibfnamefont {M.~B.}\ \bibnamefont
  {Tsang}},\ }\href {\doibase 10.1103/PhysRevC.88.044613} {\bibfield  {journal}
  {\bibinfo  {journal} {Phys. Rev. C}\ }\textbf {\bibinfo {volume} {88}},\
  \bibinfo {pages} {044613} (\bibinfo {year} {2013})}\BibitemShut {NoStop}%
\bibitem [{\citenamefont {Piantelli}\ \emph {et~al.}(2013)\citenamefont
  {Piantelli}, \citenamefont {Casini}, \citenamefont {Maurenzig}, \citenamefont
  {Olmi}, \citenamefont {Barlini}, \citenamefont {Bini}, \citenamefont
  {Carboni}, \citenamefont {Pasquali}, \citenamefont {Poggi}, \citenamefont
  {Stefanini}, \citenamefont {Valdr\`e}, \citenamefont {Bougault},
  \citenamefont {Bonnet}, \citenamefont {Borderie}, \citenamefont {Chbihi},
  \citenamefont {Frankland}, \citenamefont {Gruyer}, \citenamefont {Lopez},
  \citenamefont {Le~Neindre}, \citenamefont {P\^arlog}, \citenamefont {Rivet},
  \citenamefont {Vient}, \citenamefont {Rosato}, \citenamefont {Spadaccini},
  \citenamefont {Vigilante}, \citenamefont {Bruno}, \citenamefont {Marchi},
  \citenamefont {Morelli}, \citenamefont {Cinausero}, \citenamefont
  {Degerlier}, \citenamefont {Gramegna}, \citenamefont {Kozik}, \citenamefont
  {Twar\'og}, \citenamefont {Alba}, \citenamefont {Maiolino},\ and\
  \citenamefont {Santonocito}}]{StaggeringCasini2013}%
  \BibitemOpen
  \bibfield  {author} {\bibinfo {author} {\bibfnamefont {S.}~\bibnamefont
  {Piantelli}}, \bibinfo {author} {\bibfnamefont {G.}~\bibnamefont {Casini}},
  \bibinfo {author} {\bibfnamefont {P.~R.}\ \bibnamefont {Maurenzig}}, \bibinfo
  {author} {\bibfnamefont {A.}~\bibnamefont {Olmi}}, \bibinfo {author}
  {\bibfnamefont {S.}~\bibnamefont {Barlini}}, \bibinfo {author} {\bibfnamefont
  {M.}~\bibnamefont {Bini}}, \bibinfo {author} {\bibfnamefont {S.}~\bibnamefont
  {Carboni}}, \bibinfo {author} {\bibfnamefont {G.}~\bibnamefont {Pasquali}},
  \bibinfo {author} {\bibfnamefont {G.}~\bibnamefont {Poggi}}, \bibinfo
  {author} {\bibfnamefont {A.~A.}\ \bibnamefont {Stefanini}}, \bibinfo {author}
  {\bibfnamefont {S.}~\bibnamefont {Valdr\`e}}, \bibinfo {author}
  {\bibfnamefont {R.}~\bibnamefont {Bougault}}, \bibinfo {author}
  {\bibfnamefont {E.}~\bibnamefont {Bonnet}}, \bibinfo {author} {\bibfnamefont
  {B.}~\bibnamefont {Borderie}}, \bibinfo {author} {\bibfnamefont
  {A.}~\bibnamefont {Chbihi}}, \bibinfo {author} {\bibfnamefont {J.~D.}\
  \bibnamefont {Frankland}}, \bibinfo {author} {\bibfnamefont {D.}~\bibnamefont
  {Gruyer}}, \bibinfo {author} {\bibfnamefont {O.}~\bibnamefont {Lopez}},
  \bibinfo {author} {\bibfnamefont {N.}~\bibnamefont {Le~Neindre}}, \bibinfo
  {author} {\bibfnamefont {M.}~\bibnamefont {P\^arlog}}, \bibinfo {author}
  {\bibfnamefont {M.~F.}\ \bibnamefont {Rivet}}, \bibinfo {author}
  {\bibfnamefont {E.}~\bibnamefont {Vient}}, \bibinfo {author} {\bibfnamefont
  {E.}~\bibnamefont {Rosato}}, \bibinfo {author} {\bibfnamefont
  {G.}~\bibnamefont {Spadaccini}}, \bibinfo {author} {\bibfnamefont
  {M.}~\bibnamefont {Vigilante}}, \bibinfo {author} {\bibfnamefont
  {M.}~\bibnamefont {Bruno}}, \bibinfo {author} {\bibfnamefont
  {T.}~\bibnamefont {Marchi}}, \bibinfo {author} {\bibfnamefont
  {L.}~\bibnamefont {Morelli}}, \bibinfo {author} {\bibfnamefont
  {M.}~\bibnamefont {Cinausero}}, \bibinfo {author} {\bibfnamefont
  {M.}~\bibnamefont {Degerlier}}, \bibinfo {author} {\bibfnamefont
  {F.}~\bibnamefont {Gramegna}}, \bibinfo {author} {\bibfnamefont
  {T.}~\bibnamefont {Kozik}}, \bibinfo {author} {\bibfnamefont
  {T.}~\bibnamefont {Twar\'og}}, \bibinfo {author} {\bibfnamefont
  {R.}~\bibnamefont {Alba}}, \bibinfo {author} {\bibfnamefont {C.}~\bibnamefont
  {Maiolino}}, \ and\ \bibinfo {author} {\bibfnamefont {D.}~\bibnamefont
  {Santonocito}} (\bibinfo {collaboration} {FAZIA Collaboration}),\ }\href
  {\doibase 10.1103/PhysRevC.88.064607} {\bibfield  {journal} {\bibinfo
  {journal} {Phys. Rev. C}\ }\textbf {\bibinfo {volume} {88}},\ \bibinfo
  {pages} {064607} (\bibinfo {year} {2013})}\BibitemShut {NoStop}%
\bibitem [{\citenamefont {Kucharek}\ \emph {et~al.}(1989)\citenamefont
  {Kucharek}, \citenamefont {Ring},\ and\ \citenamefont
  {Schuck}}]{PairingRingShuck1989}%
  \BibitemOpen
  \bibfield  {author} {\bibinfo {author} {\bibfnamefont {H.}~\bibnamefont
  {Kucharek}}, \bibinfo {author} {\bibfnamefont {P.}~\bibnamefont {Ring}}, \
  and\ \bibinfo {author} {\bibfnamefont {P.}~\bibnamefont {Schuck}},\ }\href
  {\doibase 10.1007/BF01294212} {\bibfield  {journal} {\bibinfo  {journal} {Z.
  Phys. A}\ }\textbf {\bibinfo {volume} {334}},\ \bibinfo {pages} {119}
  (\bibinfo {year} {1989})}\BibitemShut {NoStop}%
\bibitem [{\citenamefont {Goriely}(1996)}]{pairingGoriely1996}%
  \BibitemOpen
  \bibfield  {author} {\bibinfo {author} {\bibfnamefont {S.}~\bibnamefont
  {Goriely}},\ }\href {\doibase http://dx.doi.org/10.1016/0375-9474(96)00162-5}
  {\bibfield  {journal} {\bibinfo  {journal} {Nuclear Physics A}\ }\textbf
  {\bibinfo {volume} {605}},\ \bibinfo {pages} {28 } (\bibinfo {year}
  {1996})}\BibitemShut {NoStop}%
\bibitem [{\citenamefont {Souza}\ \emph {et~al.}(2013)\citenamefont {Souza},
  \citenamefont {Carlson}, \citenamefont {Donangelo}, \citenamefont {Lynch},\
  and\ \citenamefont {Tsang}}]{smmde2013}%
  \BibitemOpen
  \bibfield  {author} {\bibinfo {author} {\bibfnamefont {S.~R.}\ \bibnamefont
  {Souza}}, \bibinfo {author} {\bibfnamefont {B.~V.}\ \bibnamefont {Carlson}},
  \bibinfo {author} {\bibfnamefont {R.}~\bibnamefont {Donangelo}}, \bibinfo
  {author} {\bibfnamefont {W.~G.}\ \bibnamefont {Lynch}}, \ and\ \bibinfo
  {author} {\bibfnamefont {M.~B.}\ \bibnamefont {Tsang}},\ }\href {\doibase
  10.1103/PhysRevC.88.014607} {\bibfield  {journal} {\bibinfo  {journal} {Phys.
  Rev. C}\ }\textbf {\bibinfo {volume} {88}},\ \bibinfo {pages} {014607}
  (\bibinfo {year} {2013})}\BibitemShut {NoStop}%
\bibitem [{\citenamefont {Carlson}\ \emph {et~al.}(2012)\citenamefont
  {Carlson}, \citenamefont {Donangelo}, \citenamefont {Souza}, \citenamefont
  {Lynch}, \citenamefont {Steiner},\ and\ \citenamefont {Tsang}}]{fbk2012}%
  \BibitemOpen
  \bibfield  {author} {\bibinfo {author} {\bibfnamefont {B.~V.}\ \bibnamefont
  {Carlson}}, \bibinfo {author} {\bibfnamefont {R.}~\bibnamefont {Donangelo}},
  \bibinfo {author} {\bibfnamefont {S.~R.}\ \bibnamefont {Souza}}, \bibinfo
  {author} {\bibfnamefont {W.~G.}\ \bibnamefont {Lynch}}, \bibinfo {author}
  {\bibfnamefont {A.~W.}\ \bibnamefont {Steiner}}, \ and\ \bibinfo {author}
  {\bibfnamefont {M.~B.}\ \bibnamefont {Tsang}},\ }\href {\doibase
  10.1016/j.nuclphysa.2011.12.009} {\bibfield  {journal} {\bibinfo  {journal}
  {Nuclear Physics A}\ }\textbf {\bibinfo {volume} {876}},\ \bibinfo {pages}
  {77} (\bibinfo {year} {2012})}\BibitemShut {NoStop}%
\bibitem [{\citenamefont {Bondorf}\ \emph
  {et~al.}(1985{\natexlab{a}})\citenamefont {Bondorf}, \citenamefont
  {Donangelo}, \citenamefont {Mishustin}, \citenamefont {Pethick},
  \citenamefont {Schulz},\ and\ \citenamefont {Sneppen}}]{smm1}%
  \BibitemOpen
  \bibfield  {author} {\bibinfo {author} {\bibfnamefont {J.~P.}\ \bibnamefont
  {Bondorf}}, \bibinfo {author} {\bibfnamefont {R.}~\bibnamefont {Donangelo}},
  \bibinfo {author} {\bibfnamefont {I.~N.}\ \bibnamefont {Mishustin}}, \bibinfo
  {author} {\bibfnamefont {C.}~\bibnamefont {Pethick}}, \bibinfo {author}
  {\bibfnamefont {H.}~\bibnamefont {Schulz}}, \ and\ \bibinfo {author}
  {\bibfnamefont {K.}~\bibnamefont {Sneppen}},\ }\href@noop {} {\bibfield
  {journal} {\bibinfo  {journal} {Nucl. Phys.}\ }\textbf {\bibinfo {volume}
  {A443}},\ \bibinfo {pages} {321} (\bibinfo {year}
  {1985}{\natexlab{a}})}\BibitemShut {NoStop}%
\bibitem [{\citenamefont {Bondorf}\ \emph
  {et~al.}(1985{\natexlab{b}})\citenamefont {Bondorf}, \citenamefont
  {Donangelo}, \citenamefont {Mishustin},\ and\ \citenamefont {Schulz}}]{smm2}%
  \BibitemOpen
  \bibfield  {author} {\bibinfo {author} {\bibfnamefont {J.~P.}\ \bibnamefont
  {Bondorf}}, \bibinfo {author} {\bibfnamefont {R.}~\bibnamefont {Donangelo}},
  \bibinfo {author} {\bibfnamefont {I.~N.}\ \bibnamefont {Mishustin}}, \ and\
  \bibinfo {author} {\bibfnamefont {H.}~\bibnamefont {Schulz}},\ }\href@noop {}
  {\bibfield  {journal} {\bibinfo  {journal} {Nucl. Phys.}\ }\textbf {\bibinfo
  {volume} {A444}},\ \bibinfo {pages} {460} (\bibinfo {year}
  {1985}{\natexlab{b}})}\BibitemShut {NoStop}%
\bibitem [{\citenamefont {Sneppen}(1987)}]{smm4}%
  \BibitemOpen
  \bibfield  {author} {\bibinfo {author} {\bibfnamefont {K.}~\bibnamefont
  {Sneppen}},\ }\href@noop {} {\bibfield  {journal} {\bibinfo  {journal} {Nucl.
  Phys.}\ }\textbf {\bibinfo {volume} {A470}},\ \bibinfo {pages} {213}
  (\bibinfo {year} {1987})}\BibitemShut {NoStop}%
\bibitem [{\citenamefont {Souza}\ \emph {et~al.}(2003)\citenamefont {Souza},
  \citenamefont {Danielewicz}, \citenamefont {{\protect Das Gupta}},
  \citenamefont {Donangelo}, \citenamefont {Friedman}, \citenamefont {Lynch},
  \citenamefont {Tan},\ and\ \citenamefont {Tsang}}]{ISMMmass}%
  \BibitemOpen
  \bibfield  {author} {\bibinfo {author} {\bibfnamefont {S.~R.}\ \bibnamefont
  {Souza}}, \bibinfo {author} {\bibfnamefont {P.}~\bibnamefont {Danielewicz}},
  \bibinfo {author} {\bibfnamefont {S.}~\bibnamefont {{\protect Das Gupta}}},
  \bibinfo {author} {\bibfnamefont {R.}~\bibnamefont {Donangelo}}, \bibinfo
  {author} {\bibfnamefont {W.~A.}\ \bibnamefont {Friedman}}, \bibinfo {author}
  {\bibfnamefont {W.~G.}\ \bibnamefont {Lynch}}, \bibinfo {author}
  {\bibfnamefont {W.~P.}\ \bibnamefont {Tan}}, \ and\ \bibinfo {author}
  {\bibfnamefont {M.~B.}\ \bibnamefont {Tsang}},\ }\href@noop {} {\bibfield
  {journal} {\bibinfo  {journal} {Phys. Rev. C}\ }\textbf {\bibinfo {volume}
  {67}},\ \bibinfo {pages} {051602(R)} (\bibinfo {year} {2003})}\BibitemShut
  {NoStop}%
\bibitem [{\citenamefont {Wigner}\ and\ \citenamefont
  {Seitz}(1934)}]{WignerSeitz}%
  \BibitemOpen
  \bibfield  {author} {\bibinfo {author} {\bibfnamefont {E.}~\bibnamefont
  {Wigner}}\ and\ \bibinfo {author} {\bibfnamefont {F.}~\bibnamefont {Seitz}},\
  }\href@noop {} {\bibfield  {journal} {\bibinfo  {journal} {Phys. Rev.}\
  }\textbf {\bibinfo {volume} {46}},\ \bibinfo {pages} {509} (\bibinfo {year}
  {1934})}\BibitemShut {NoStop}%
\bibitem [{\citenamefont {Pratt}\ and\ \citenamefont
  {Das~Gupta}(2000)}]{PrattDasGupta2000}%
  \BibitemOpen
  \bibfield  {author} {\bibinfo {author} {\bibfnamefont {S.}~\bibnamefont
  {Pratt}}\ and\ \bibinfo {author} {\bibfnamefont {S.}~\bibnamefont
  {Das~Gupta}},\ }\href {\doibase 10.1103/PhysRevC.62.044603} {\bibfield
  {journal} {\bibinfo  {journal} {Phys. Rev. C}\ }\textbf {\bibinfo {volume}
  {62}},\ \bibinfo {pages} {044603} (\bibinfo {year} {2000})}\BibitemShut
  {NoStop}%
\bibitem [{\citenamefont {Tan}\ \emph {et~al.}(2003)\citenamefont {Tan},
  \citenamefont {Souza}, \citenamefont {Charity}, \citenamefont {Donangelo},
  \citenamefont {Lynch},\ and\ \citenamefont {Tsang}}]{ISMMlong}%
  \BibitemOpen
  \bibfield  {author} {\bibinfo {author} {\bibfnamefont {W.~P.}\ \bibnamefont
  {Tan}}, \bibinfo {author} {\bibfnamefont {S.~R.}\ \bibnamefont {Souza}},
  \bibinfo {author} {\bibfnamefont {R.~J.}\ \bibnamefont {Charity}}, \bibinfo
  {author} {\bibfnamefont {R.}~\bibnamefont {Donangelo}}, \bibinfo {author}
  {\bibfnamefont {W.~G.}\ \bibnamefont {Lynch}}, \ and\ \bibinfo {author}
  {\bibfnamefont {M.~B.}\ \bibnamefont {Tsang}},\ }\href@noop {} {\bibfield
  {journal} {\bibinfo  {journal} {Phys. Rev. C}\ }\textbf {\bibinfo {volume}
  {68}},\ \bibinfo {pages} {034609} (\bibinfo {year} {2003})}\BibitemShut
  {NoStop}%
\bibitem [{\citenamefont {Gilbert}\ and\ \citenamefont
  {Cameron}(1965)}]{levelDensityGilbertCameron1965}%
  \BibitemOpen
  \bibfield  {author} {\bibinfo {author} {\bibfnamefont {A.}~\bibnamefont
  {Gilbert}}\ and\ \bibinfo {author} {\bibfnamefont {A.~G.~W.}\ \bibnamefont
  {Cameron}},\ }\href {\doibase 10.1139/p65-139} {\bibfield  {journal}
  {\bibinfo  {journal} {Canadian Journal of Physics}\ }\textbf {\bibinfo
  {volume} {43}},\ \bibinfo {pages} {1446} (\bibinfo {year} {1965})},\ \Eprint
  {http://arxiv.org/abs/http://dx.doi.org/10.1139/p65-139}
  {http://dx.doi.org/10.1139/p65-139} \BibitemShut {NoStop}%
\bibitem [{\citenamefont {Ring}\ and\ \citenamefont {Schuck}(1980)}]{rs80}%
  \BibitemOpen
  \bibfield  {author} {\bibinfo {author} {\bibfnamefont {P.}~\bibnamefont
  {Ring}}\ and\ \bibinfo {author} {\bibfnamefont {P.}~\bibnamefont {Schuck}},\
  }\href@noop {} {\emph {\bibinfo {title} {The nuclear many-body problem}}}\
  (\bibinfo  {publisher} {Springer-Verlag, New York, Heidelberg, and Berlin},\
  \bibinfo {year} {1980})\BibitemShut {NoStop}%
\end{thebibliography}%
\bibliographystyle{apsrev4-1}

\end{document}